\documentclass[lettersize,journal]{IEEEtran}
\usepackage{amsmath,amsfonts}
\usepackage{algorithmic}
\usepackage{algorithm}
\usepackage{amssymb}
\usepackage{array}
\usepackage{textcomp}
\usepackage{stfloats}
\usepackage{url}
\usepackage{verbatim}
\usepackage{graphicx}
\usepackage{cite}
\usepackage{subfigure}
\usepackage{color}
\usepackage{mathtools}
\usepackage{float}
\usepackage{caption}
\usepackage{xcolor}
\usepackage{svg}
\usepackage{mathtools}
\begin{document}
		\title{Rotatable Antenna-Enhanced Secure Integrated Sensing and Communications Under Imperfect CSI}
		\author{Qi~Yang,~\IEEEmembership{Graduate Student Member,~IEEE,}
			Kai~Liu,~\IEEEmembership{Member,~IEEE,}
			Jingjing~Zhao,~\IEEEmembership{Senior Member,~IEEE,}
			Xidong~Mu,~\IEEEmembership{Member,~IEEE,}
			Tianqi~Mao,~\IEEEmembership{Member,~IEEE,}
			and Kaiquan~Cai,~\IEEEmembership{Member,~IEEE}
			
			\vspace{-5mm}
			
			\thanks{Q. Yang, K. Liu, J. Zhao, and K. Cai are with the School of Electronics and Information Engineering, Beihang University, Beijing, 100191 China, and also with the State Key Laboratory of CNS/ATM, Beijing 100191, China. (e-mail:\{yangqi01, liuk, jingjingzhao, caikq\}@buaa.edu.cn). }
			
			\thanks{X. Mu is with the Centre for Wireless Innovation (CWI), Queen's University Belfast, Belfast, BT3 9DT, U.K. (e-mail: x.mu@qub.ac.uk). }
			
			\thanks{T. Mao is with the State Key Laboratory of Environment Characteristics and Effects for Near-Space, Beijing Institute of Technology, Beijing 100081, China, and also with Beijing Institute of Technology, Zhuhai 519088, China (e-mail: maotq@bit.edu.cn). }}

		\maketitle
	
	\begin{abstract}
		A rotatable antenna (RA)-enhanced secure integrated sensing and communications system is investigated, where an RA-based transceiver simultaneously communicates with legitimate users and senses a target that is regarded as a potential eavesdropper.
		Under imperfect eavesdropping channel state information (CSI), a max-min data rate optimization problem is formulated by jointly optimizing the transmit beamforming, artificial noise (AN) covariance matrix, and transmit/receive boresights of RAs, subject to the maximum information leakage and minimum sensing power constraints.
		To address the highly non-convex problem, the information leakage and sensing power constraints are transformed into convex ones via S-Procedure method and Cauchy-Schwarz inequality, respectively.
		Subsequently, an alternating optimization algorithm is developed to decompose the reformulated problem into two subproblems.
		In particular, the transmit beamforming and AN covariance matrix are optimized by utilizing successive convex approximation and semi-definite relaxation methods, while the RA boresights are obtained by invoking the particle swarm optimization.
		Simulation results show that the RA-based scheme significantly outperforms the benchmarks, and offers enhanced robustness against imperfect CSI with the increase of the maximum rotation range.
	\end{abstract}
		
	
	\section{Introduction}
	
	\IEEEPARstart{I}{ntegrated} sensing and communications (ISAC) has emerged as a promising technology for future wireless networks, owing to its ability to support sensing and communication functionalities within a unified framework~\cite{Zhang2026}.
	Specifically, ISAC enables simultaneous sensing and communication by sharing resources such as waveform, spectrum, and hardware platform, thereby enhancing spectral efficiency and reducing deployment costs.
	However, the shared waveform may compromise transmission security, as the information intended for legitimate users is inevitably leaked to sensing targets that might act as potential eavesdroppers (Eves)~\cite{ZhaoNear2026}.
	To address this issue, physical layer security (PLS) offers a viable solution by exploiting the spatial channel disparity between legitimate users and Eves, without relying on complex cryptographic designs.
	In particular, the transmitter can ensure secure transmission by employing PLS techniques, such as secure beamforming~\cite{LiaoRobust2023} and artificial noise (AN)~\cite{WangArtificial2025}.
	Nevertheless, these techniques are mostly applied in conventional antenna systems with fixed positions and orientations, which lack sufficient degrees of freedom (DoFs) to adapt to spatial channel variations.
	
	Recently, movable antennas (MAs)~\cite{XiaoChannel2024} and fluid antennas (FAs)~\cite{LuFluid2025} have been proposed to enhance communication performance through flexible antenna position adjustments.
	However, despite their potential, MAs and FAs are constrained by limited response speed and isotropic radiation patterns, which restrict their practical applicability in dynamic environments.
	Motivated by this, rotatable antennas (RAs) have garnered increasing attention due to their ability to reshape radiation patterns while maintaining a compact aperture and low hardware complexity~\cite{ZhengRotatable2025Opportunities}.
	By independently adjusting the three-dimensional (3D) boresight of each antenna via mechanical or electronic methods, RAs can provide additional spatial DoFs to enhance the array gain towards users, while suppressing radiation power in undesired directions \cite{XiongIntelligent2025}.
	Unlike MAs and FAs that require extra physical space for antenna movement, RAs only need local rotation adjustments of antenna boresights, making it highly scalable and compatible with existing wireless infrastructures.
	Driven by these distinct advantages, recent studies explored the performance of RAs in various scenarios, including ISAC~\cite{ZhouRotatable2025}, PLS~\cite{DaiRotatable2025}, and cognitive radio systems~\cite{TanRotatable2026Cognitive}.
	However, the potential of RAs for enhancing secure ISAC performance remains unexplored, especially with the consideration of imperfect eavesdropping channel state information (CSI).
	
	Motivated by the above, we consider a RA-enhanced secure ISAC system, where an RA-based transceiver is deployed to communicate with legitimate users and sense an Eve with imperfect CSI.
	To guarantee user fairness, we formulate a max-min rate optimization problem for all users by jointly optimizing the transmit beamforming, AN covariance matrix, and transmit/receive boresights.
	To address the intractable problem, we apply the S-Procedure and Cauchy-Schwarz inequality to convexify the semi-infinite constraints induced by the imperfect eavesdropping CSI.
	Subsequently, we develop an alternating optimization (AO) algorithm, where the transmit beamforming and AN covariance matrix are optimized via successive convex approximation (SCA) and semi-definite relaxation (SDR), while the boresights are obtained by invoking the particle swarm optimization (PSO).
	Simulation results demonstrate that the proposed RA-based scheme significantly outperforms the benchmarks with fixed-orientation or isotropic antennas in both sensing and communication performance.
	
	\section{System Model and Problem Formulation}
	\begin{figure}[t]
		\centering
		\includegraphics[width=3in]{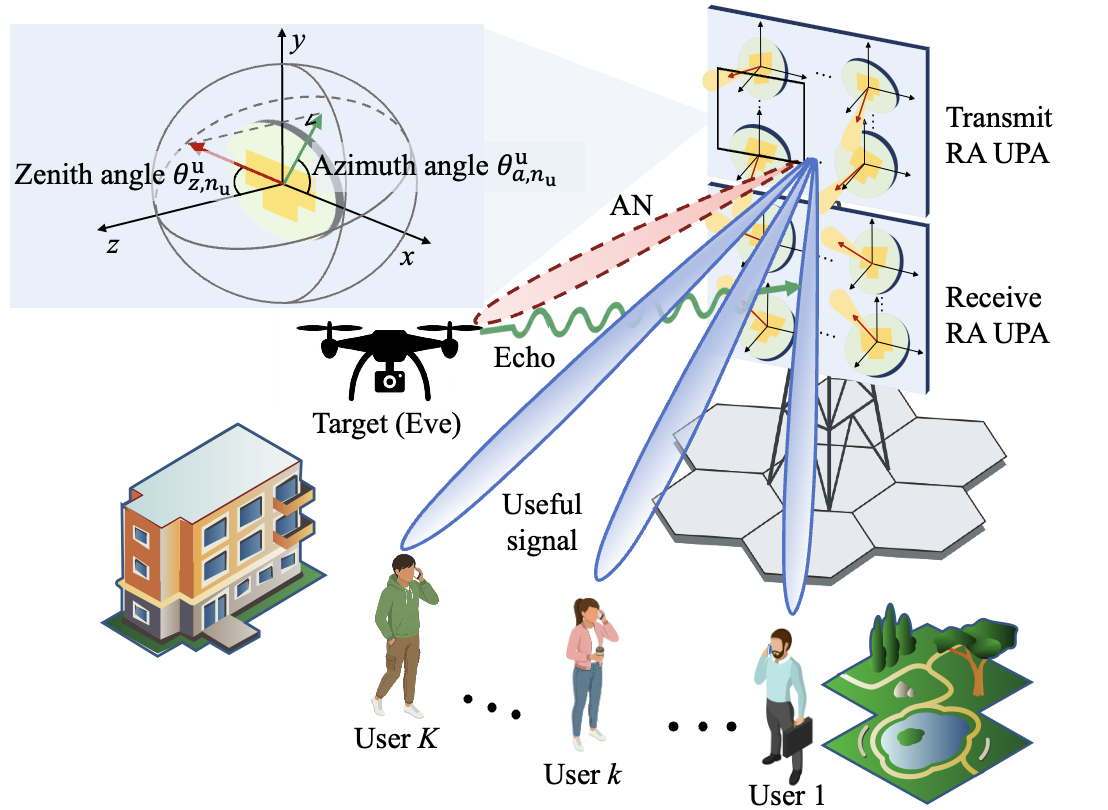}
		\caption{Illustration of the RA-enhanced secure ISAC.}
		\label{rotateble_scenario1}
		\vspace{-3mm}
	\end{figure}
	As illustrated in Fig. \ref{rotateble_scenario1}, an RA-enhanced secure ISAC system is considered, where a transceiver equipped with two RA-based uniform planar arrays (UPAs) communicates with $K$ single-antenna users, while simultaneously sensing a single-antenna Eve.
	The transmit and receive UPAs consist of $N_{\text{t}}$ and $N_{\text{r}}$ antennas, respectively. We assume that $N_{\text{t}}=N_{\text{r}}=N=N_{\text{h}} \times N_{\text{v}}$, where $N_{\text{h}}$ and $N_{\text{v}}$ denote the number of RAs in the horizontal and vertical dimensions, respectively. 
    In a 3D Cartesian coordinate system where the UPAs are deployed in the $x$-$y$ plane, the position vectors of the Eve and the $k$-th user are denoted by $\mathbf{p}_0$ and $\mathbf{p}_k$, $k \geq 1$, respectively.
	 \vspace{-3mm}
	\subsection{RA Model}
	
	The boresight vector of the $n_\text{u}$-th RA is denoted by $\boldsymbol{\theta}^\text{u}_{n_\text{u}} = \left[\theta^\text{u}_{z, n_\text{u}}, \theta^\text{u}_{a, n_\text{u}}\right]^T$, where $\text{u} \in \left\{\text{t}, \text{r}\right\}$ represents the transmit/receive UPA, $\theta^\text{u}_{z, n_\text{u}}$ is the zenith angle between the RA's boresight and the $z$-axis, and $\theta^\text{u}_{a, n_\text{u}}$ represents the azimuth angle between the boresight's projection onto $x$-$y$ plane and the $x$-axis.
	Accordingly, the pointing vector of the $n_\text{u}$-th RA is given by 
	$\mathbf{r}(\boldsymbol{\theta}^\text{u}_{n_\text{u}}) = \left[\sin \theta^\text{u}_{z, n_\text{u}} \cos \theta^\text{u}_{a, n_\text{u}}, \sin \theta^\text{u}_{z, n_\text{u}} \sin \theta^\text{u}_{a, n_\text{u}}, \cos \theta^\text{u}_{z, n_\text{u}} \right]^T$.
	
	Following the cosine pattern model provided in \cite{balanis2016antenna}, the directional gain of an RA is given by
	\begin{equation}
		G(\epsilon) = 
		\begin{cases} 
			G_0 \cos^{2p}(\epsilon), & \epsilon \in [0, \frac{\pi}{2}), \\ 
			0, & \text{otherwise},
		\end{cases}
	\end{equation}
    where $\epsilon$ denotes the angular offset between any spatial direction and the RA's boresight, $p$ represents the directivity factor, and $G_0 = 2(2p + 1)$ is the maximum directional gain at $\epsilon = 0$.
	Accordingly, the directional gain of the $n_\text{u}$-th RA towards the $k$-th user/Eve is $G_{ k}(\boldsymbol{\theta}^\text{u}_{n_\text{u}})= G_0 \cos^{2p}(\epsilon^\text{u}_{n_\text{u}, k})$, where $\cos(\epsilon^\text{u}_{n_\text{u}, k}) = \mathbf{r}^T(\boldsymbol{\theta}^\text{u}_{n_\text{u}}) \frac{ \mathbf{p}_k}{\|\mathbf{p}_k\|}$ is the projection of the pointing vector of the $n_\text{u}$-th RA onto the direction vector of the $k$-th user/Eve, where $\| \cdot \|$ is the $l_2$-norm of a vector.

	\vspace{-3mm}
	\subsection{Channel and Signal Model}
	
	Given that the propagation distances between the RA-based transceiver and users/Eve are much larger than the apertures of UPAs, the propagation channels are characterized by the far-field model with planar wavefronts.
	Consequently, the array response vector for the $k$-th user/Eve is given by
	\begin{equation}
		\begin{aligned}
			\mathbf{a}_k= & \left[1, e^{-j\frac{2\pi d}{\lambda}\alpha_k}, ... , e^{-j \frac{2\pi d}{\lambda}(N_{\text{h}}-1)\alpha_k}\right]^T\\
			& \otimes \left[1, e^{-j\frac{2\pi d}{\lambda}\beta_k}, ... , e^{-j \frac{2\pi d}{\lambda}(N_{\text{v}}-1)\beta_k}\right]^T,
		\end{aligned}
	\end{equation}
	where $k\in\left\{0, ..., K\right\}$, $d$ denotes the antenna spacing, $\lambda$ is the signal wavelength, $\alpha_k = \sin \phi_k \cos \psi_k$ and $\beta_k = \cos \phi_k$.
	Here, $\phi_k$ and $\psi_k$ represent the elevation and azimuth angles of the angle-of-arrival (AoA)/angle-of-departure (AoD) for the $k$-th user/Eve, respectively.
	Moreover, the RA radiation coefficient vector for the $k$-th user/Eve is given by
	\begin{equation}
		\mathbf{b}_k\left(\boldsymbol{\Theta}_{\text{u}}\right) = \left[ \sqrt{G_{k}\left(\boldsymbol{\theta}^\text{u}_{1}\right)}, \dots, \sqrt{G_{k}\left(\boldsymbol{\theta}^\text{u}_{N_\text{u}}\right)} \right]^T,
	\end{equation}
	where $\boldsymbol{\Theta}_{\text{u}} = \left[\boldsymbol{\theta}^\text{u}_{1}, ... , \boldsymbol{\theta}^\text{u}_{N_\text{u}}\right]$ denotes the boresight matrix of the transmit/receive UPA.
    Under the Rician channel model, the channel vector between the transmit UPA and user $k$ is given by
    \begin{equation}
    	\label{rician}
    	\mathbf{h}_k(\mathbf{\Theta}_\text{t}) = \rho_k \mathbf{b}_k \left(\boldsymbol{\Theta}_{\text{t}}\right)  \odot \left( \sqrt{\frac{\omega_k}{\omega_k + 1}} \mathbf{a}_k + \sqrt{\frac{1}{\omega_k + 1}} \mathbf{g}_{k} \right),
    \end{equation}
    where $\rho_k = \sqrt{\zeta r_k^{-\ell_u}}$ denotes the complex path loss between the transmitter and user $k$,  $\zeta$ is the path loss at the reference distance $1$~m, $r_k = \|\mathbf{p}_k\|$, $\ell_u$ is the path loss exponent for users. Still referring to \eqref{rician}, $\omega_k$ is the Rician factor, and $\mathbf{g}_{k} \sim \mathcal{CN}(\mathbf{0}, \mathbf{I}_{N})$ represents the non line-of-sight (NLoS) component.
    Moreover, we consider a line-of-sight (LoS) channel between the transmitter/receiver and the Eve, which can be expressed as $\mathbf{h}_0\left(\boldsymbol{\Theta}_{\text{u}}\right) = \sqrt{\zeta r_0^{-\ell_e}} \mathbf{b}_0 \left(\boldsymbol{\Theta}_{\text{u}}\right)  \odot \mathbf{a}_0$, where $r_0 = \|\mathbf{p}_0\|$, and $\ell_e$  is the path loss exponent for the Eve.
    Then, we have  the round-trip sensing channel given by $\mathbf{H}_{\text{R}} = \eta \mathbf{h}_0\left(\boldsymbol{\Theta}_{{\text{r}}}\right) \mathbf{h}^H_0\left(\boldsymbol{\Theta}_{\text{t}}\right) \in \mathbb{C}^{N_{\text{r}} \times N_{\text{t}}}$, where $\eta$ denotes the complex reflection coefficient accounting for the Eve's radar cross section (RCS).
	
	Considering that the Eve's perfect CSI is normally unavailable at the transceiver, we adopt the deterministic bounded error model for both the eavesdropping and round-trip sensing channels, which are given by
	\begin{subequations}
		\begin{align}
			\label{csi_h_0}
			&\mathbf{h}_0\left(\boldsymbol{\Theta}_{\text{t}}\right) = 
			\hat{\mathbf{h}}_0\left(\boldsymbol{\Theta}_{\text{t}}\right) + \Delta \mathbf{h}, \|\Delta \mathbf{h}\| \leq \varepsilon_{h}, \\
			\label{csi_H_R}
			&\mathbf{H}_{\text{R}} =
			\hat{\mathbf{H}}_{\text{R}}\ + \Delta \mathbf{H}_{\text{R}}, \|\Delta \mathbf{H}_{\text{R}}\|_F \leq \varepsilon_H,
		\end{align}
	\end{subequations}
	where $\|\cdot\|_F$ is the Frobenius norm, $\hat{\mathbf{h}}_0\left(\boldsymbol{\Theta}_{\text{t}}\right)$ and $\hat{\mathbf{H}}_{\text{R}}$ denote the estimated channels, $\Delta \mathbf{\mathbf{h}}$ and $\Delta \mathbf{H}_{\text{R}}$ are the CSI errors, with the maximum thresholds $\varepsilon_h$ and $\varepsilon_H$, respectively.
	
	The signal $\mathbf{x}$ transmitted by the transceiver is composed of the information symbols and AN, which is given by
	\begin{equation}
		\mathbf{x} = \sum_{k=1}^K \mathbf{w}_k c_k +  \mathbf{s}_{\text{AN}},
	\end{equation}
	where $\mathbf{w}_k \in \mathbb{C}^{N_{\text{t}} \times 1}$ and $c_k$ represent the transmit beamforming vector and the information symbol for the $k$-th user, respectively.
	$\mathbf{s}_{\text{AN}} \in \mathbb{C}^{N_{\text{t}} \times 1}$ denotes the AN that also serves as the dedicated sensing waveform.
	The received signal at the $k$-th user/Eve is given by
	\begin{equation}
		y_k =  \mathbf{h}^{H}_{k}\left(\boldsymbol{\Theta}_{\text{t}}\right) \mathbf{x} + n_k, \forall k\in\left\{0, ..., K\right\},
	\end{equation}
	where $n_k \sim \mathcal{CN}(0, \sigma_k^2)$ is the additive white Gaussian noise (AWGN) with noise power $\sigma_k^2$.
	Moreover, the echo signal obtained at the receiver is given by
	\begin{equation}
		\mathbf{y}_e = \mathbf{H}_{\text{R}} \mathbf{x} + \mathbf{n}_e,
	\end{equation}
	where $ \mathbf{n}_e \sim \mathcal{CN}(\mathbf{0}, \sigma_e^2\mathbf{I}_{N_\textbf{r}})$ is the AWGN with power $\sigma_e^2$.
	
	The SINR at the $k$-th user and the corresponding eavesdropping SINR at the Eve are respectively given by
	\begin{equation}
		\gamma_k = \frac{\mathbf{h}^H_{k}\left(\boldsymbol{\Theta}_{\text{t}}\right) \mathbf{W}_{k} \mathbf{h}_{k}\left(\boldsymbol{\Theta}_{\text{t}}\right)}{\mathbf{h}^H_{k}\left(\boldsymbol{\Theta}_{\text{t}}\right) \left(\sum^{K}_{i = 1, i \neq k} \mathbf{W}_{i}  + \mathbf{R}_{\text{AN}}\right) \mathbf{h}_{k}\left(\boldsymbol{\Theta}_{\text{t}}\right) + \sigma^2_{k} },
	\end{equation}
	\begin{equation}
		\gamma_{e,k} = \frac{\mathbf{h}^H_{0}\left(\boldsymbol{\Theta}_{\text{t}}\right) \mathbf{W}_{k}\mathbf{h}_{0}\left(\boldsymbol{\Theta}_{\text{t}}\right)}{\mathbf{h}^H_{0}\left(\boldsymbol{\Theta}_{\text{t}}\right)\left(\sum^{K}_{i = 1, i \neq k}  \mathbf{W}_{i} + \mathbf{R}_{\text{AN}}\right)\mathbf{h}_{0}\left(\boldsymbol{\Theta}_{\text{t}}\right) + \sigma^2_{0} },
	\end{equation}
	where $\mathbf{W}_k \triangleq \mathbf{w}_k \mathbf{w}^H_k$ satisfies $\mathbf{W}_k \succeq 0$ and $\text{rank}\left(\mathbf{W}_k\right) = 1$, $\mathbf{R}_{\text{AN}} = \mathbb{E}[\mathbf{s}_{\text{AN}} \mathbf{s}_{\text{AN}}^H]$ is the AN covariance matrix.
	Accordingly, the achievable rates are given by $R_k = \log_2 \left(1+\gamma_k\right)$ and $R_{e,k} = \log_2 \left(1+\gamma_{e,k}\right)$, respectively. 
	
	\subsection{Problem Formulation}
	
	In this paper, we aim to maximize the minimum achievable rate for all users by jointly optimizing the transmit beamforming, AN covariance matrix, and transmit/receive boresights, subject to both the information leakage and sensing power constraints. 
	The optimization problem can be formulated as follows:
	\begin{subequations}
		\label{eq:optimization_problem}
		\begin{align}
			\label{eq:objective_function}
			& \max_{\{\mathbf{W}_k\}, \mathbf{R}_{\text{AN}}, \boldsymbol{\Theta}_{\text{t}}, \boldsymbol{\Theta}_{\text{r}}} \min_{k} R_k, \\
			\label{eq:constraint_eve_rate}
			\text{s.t.} \quad & \max_{\|\Delta \mathbf{h}\| \leq \varepsilon_h} R_{e,k} \leq \xi_k, \forall k, \\
			\label{eq:constraint_sen_power}
			& \min_{\|\Delta \mathbf{H}_{\text{R}}\|_F \leq \varepsilon_H} \text{Tr}\left(\mathbf{H}_{\text{R}} \mathbf{R}_{\text{x}} \mathbf{H}^H_{\text{R}}\right) \geq \tau, \\
			\label{eq:constraint_theta_u}
			& 0 \leq \theta^{\text{u}}_{z,n_{\text{u}}} \leq \theta_{\max}, \forall n_{\text{u}}, \\
			\label{eq:constraint_trans_power}
			& \sum_{k=1}^{K}\text{Tr}(\mathbf{W}_k) + \text{Tr}\left(\mathbf{R}_{\text{AN}}\right) \leq P_{\max}, \\
			\label{eq:constraint_sdr}
			& \mathbf{W}_k \succeq 0, \mathbf{R}_{\text{AN}} \succeq 0, \forall k, \\
			\label{eq:constraint_rank1}
			& \text{rank}(\mathbf{W}_k) = 1, \forall k,
		\end{align}
	\end{subequations}
	where $\mathbf{R}_{\text{x}} = \sum_{k=1}^{K} \mathbf{W}_k + \mathbf{R}_{\text{AN}}$ denotes the transmit covariance matrix. \eqref{eq:constraint_eve_rate} gives the constraint on the maximum information leakage for the $k$-th user's signal with the threshold $\xi_k$. \eqref{eq:constraint_sen_power} ensures the sensing power should be larger than the threshold $\tau$. 
	\eqref{eq:constraint_theta_u} restricts the maximum rotation range $\theta_{\max}$ for each RA's zenith angle.
	\eqref{eq:constraint_trans_power} gives the maximum transmit power budget $P_{\max}$. 
	Moreover, \eqref{eq:constraint_sdr} and \eqref{eq:constraint_rank1} ensure that $\mathbf{W}_k$ and $\mathbf{R}_{\text{AN}}$ are positive semidefinite matrices, and $\mathbf{W}_k$ satisfies the rank-one constraint.
	Problem \eqref{eq:optimization_problem} is challenging to solve due to its non-convexity, coupled variables, and the semi-infinite constraints induced by the CSI uncertainty.
	 
	 \vspace{-3mm}
	\section{Proposed Solution}
	
	In this section, we first transform problem \eqref{eq:optimization_problem} into a more tractable one by implementing the S-Procedure and Cauchy-Schwarz inequality, and then employ the AO algorithm to iteratively solve two subproblems.
	
	\subsection{Problem Reformulation}
	For addressing the semi-infinite constraint \eqref{eq:constraint_eve_rate}, we first reformulate it into an equivalent SINR form, i.e., $\max_{\|\Delta \mathbf{h}\| \leq \varepsilon_h} \gamma_{e,k} \leq \delta_k$, where $\delta_k = 2^{\xi_k} - 1$.
	For simplicity of illustration, let $\hat{\mathbf{h}}_0$ denote $\hat{\mathbf{h}}_0\left(\boldsymbol{\Theta}_{\text{t}}\right)$.
	By substituting \eqref{csi_h_0} into the inequality, we can obtain the following constraint:
	\begin{equation}
		\label{eq:e_k_constrant}
		\hat{\mathbf{h}}^H_0 \mathbf{P}_k \hat{\mathbf{h}}_0 + 2 \text{Re}\{\hat{\mathbf{h}}^H_0 \mathbf{P}_k \Delta \mathbf{h}\} + \Delta \mathbf{h}^H \mathbf{P}_k \Delta \mathbf{h} + \delta_k\sigma^2_0 \geq 0,
	\end{equation} 
	where $\mathbf{P}_k = \delta_k \sum_{i=1, i \neq k}^{K} \mathbf{W}_i + \delta_k \mathbf{R}_{\text{AN}} - \mathbf{W}_k$.
	According to the S-Procedure \cite{LiaoRobust2023}, the inequality \eqref{eq:e_k_constrant} is equivalent to the following linear matrix inequality:
	\begin{equation}
		\label{eq:s_procedure_constrant}
		\mathbf{A}_k = 
		\begin{bmatrix}
			\mathbf{P}_k + \lambda_k \mathbf{I} & \mathbf{P}_k \hat{\mathbf{h}}_0\\
			\hat{\mathbf{h}}_0^H \mathbf{P}_k & \hat{\mathbf{h}}_0^H \mathbf{P}_k \hat{\mathbf{h}}_0 + \delta_k \sigma_0^2 - \lambda_k \varepsilon_h^2
		\end{bmatrix} \succeq 0,
	\end{equation}
	where $\lambda_k$ is the nonnegative auxiliary variable to be optimized later.
	
	For addressing constraint \eqref{eq:constraint_sen_power}, we substitute \eqref{csi_H_R} into it, which allows the sensing power to be equivalently rewritten as
	\begin{align}
		\label{eq:trace_expansion}
		\text{Tr}(\mathbf{H}_{\text{R}} \mathbf{R}_{\text{x}} \mathbf{H}^H_{\text{R}}) &= 
		\text{Tr}(\hat{\mathbf{H}}_{\text{R}} \mathbf{R}_{\text{x}} \hat{\mathbf{H}}^H_{\text{R}}) + 2 \text{Re}\{ \text{Tr}( \hat{\mathbf{H}}_{\text{R}} \mathbf{R}_{\text{x}} \Delta \mathbf{H}^H_{\text{R}} ) \} \notag \\
		&\quad + \text{Tr}(\Delta \mathbf{H}_{\text{R}} \mathbf{R}_{\text{x}} \Delta \mathbf{H}^H_{\text{R}}).
	\end{align}
	By applying the Cauchy-Schwarz inequality with $\|\Delta \mathbf{H}_{\text{R}}\|_F \leq \varepsilon_H$, the lower bound of the term $2 \text{Re}\{ \text{Tr}( \hat{\mathbf{H}}_{\text{R}} \mathbf{R}_{\text{x}} \Delta \mathbf{H}^H_{\text{R}} ) \}$ can be obtained as follows:
	\begin{equation}
		\begin{aligned}
			2 \text{Re}\{ \text{Tr}( \hat{\mathbf{H}}_{\text{R}} \mathbf{R}_{\text{x}} \Delta \mathbf{H}^H_{\text{R}} ) \}
			&\geq -2 \| \mathbf{R}_{\text{x}} \hat{\mathbf{H}}^H_{\text{R}} \|_F \| \Delta \mathbf{H}_{\text{R}} \|_F\\
			&\geq - 2 \varepsilon_H \| \mathbf{R}_{\text{x}} \hat{\mathbf{H}}^H_{\text{R}} \|_F.
		\end{aligned}
	\end{equation}
	Given that $\text{Tr}(\Delta \mathbf{H}_{\text{R}} \mathbf{R}_{\text{x}} \Delta \mathbf{H}^H_{\text{R}}) \geq 0$, we can transform \eqref{eq:constraint_sen_power} into the following tractable one:
	\begin{equation}
		\label{eq:constraint_sen_power_lower_trans}
		\text{Tr}(\hat{\mathbf{H}}_{\text{R}} \mathbf{R}_{\text{x}} \hat{\mathbf{H}}^H_{\text{R}}) - 2 \varepsilon_H \| \mathbf{R}_{\text{x}} \hat{\mathbf{H}}^H_{\text{R}} \|_F \geq \tau.
	\end{equation}
	
	Consequently, the original optimization problem \eqref{eq:optimization_problem} can be reformulated as
	\begin{subequations}
		\label{eq:optimization_problem2}
		\begin{equation}
			\label{eq:objective_function2} 
			\max_{\{\mathbf{W}_k\}, \mathbf{R}_{\text{AN}}, \boldsymbol{\Theta}_{\text{t}}, \boldsymbol{\Theta}_{\text{r}}, \{\lambda_k\}} \min_{k} R_k,
		\end{equation}
		\begin{equation}
			{\rm{s.t.}} \  \   \eqref{eq:constraint_theta_u}, 	\eqref{eq:constraint_trans_power},	\eqref{eq:constraint_sdr}, \eqref{eq:constraint_rank1}, \eqref{eq:s_procedure_constrant}, \eqref{eq:constraint_sen_power_lower_trans}.
		\end{equation}
	\end{subequations}
	The problem \eqref{eq:optimization_problem2} remains non-convex, with the variables being highly coupled.
	Therefore, we propose an AO algorithm to iteratively optimize $\{\{\mathbf{W}_k\}, \mathbf{R}_{\text{AN}}, \{\lambda_k\}\}$ and  $\{\boldsymbol{\Theta}_{\text{t}},\boldsymbol{\Theta}_{\text{r}}\}$.
	
	\subsection{Transmit Beamforming and AN Matrix Optimization}
	
	For fixed $\{\boldsymbol{\Theta}_{\text{t}},\boldsymbol{\Theta}_{\text{r}}\}$, we first introduce an auxiliary variable $\Gamma$ to transform the objective function into $\max_{\{\mathbf{W}_k\}, \mathbf{R}_{\text{AN}}, \{\lambda_k\}, \Gamma} \Gamma$, s.t. $R_k \geq \Gamma, \forall k$.
	Since $R_k$ is still non-convex with respect to the optimization variables, we employ the SCA method to address it.
	Specifically, by applying the first-order Taylor expansion, a concave lower bound for $R_k$ at the $l$-th SCA iteration is derived as \eqref{eq:r_sca}, shown at the top of the next page, where 
	$C^{(l)}_k \triangleq \mathbf{h}^H_k(\boldsymbol{\Theta}_{\text{t}}) (\sum_{i \neq k} \mathbf{W}^{(l)}_i + \mathbf{R}^{(l)}_{\text{AN}}) \mathbf{h}_k(\boldsymbol{\Theta}_{\text{t}}) + \sigma_k^2$.
	\begin{figure*}
		\begin{equation}
			\label{eq:r_sca}
			R_{k} \geq \overline{R}_{k} =  \log_2 \left( \mathbf{h}^H_k(\boldsymbol{\Theta}_{\text{t}}) \mathbf{R}_{\text{x}} \mathbf{h}_k(\boldsymbol{\Theta}_{\text{t}}) + \sigma_k^2 \right) - \frac{1}{ C^{(l)}_k \ln 2} \left( \mathbf{h}^H_k(\boldsymbol{\Theta}_{\text{t}}) \left(\sum_{i \neq k} \mathbf{W}_i + \mathbf{R}_{\text{AN}}\right) \mathbf{h}_k(\boldsymbol{\Theta}_{\text{t}}) + \sigma^2_k\right) - \log_2C^{(l)}_k + \frac{1}{\ln 2}.
		\end{equation}
		\hrulefill
	\end{figure*}
	 
	 To facilitate the use of standard convex solvers, we relax the rank-one constraint  \eqref{eq:constraint_rank1} via SDR method.
	 Consequently, the optimization subproblem is reformulated as
	 \begin{subequations}
	 	\label{eq:optimization_problem4}
	 	\setlength{\jot}{1pt}
	 	\begin{align}
	 		\label{eq:objective_function4} 
	 		& \max_{\{\mathbf{W}_k\}, \mathbf{R}_{\text{AN}}, \{\lambda_k\}, \Gamma} \Gamma, \\
	 		\label{eq:r_k_min}
	 		\text{s.t.} \quad & \overline{R}_k \geq \Gamma, \forall k, \\
	 		& \eqref{eq:constraint_trans_power}, \eqref{eq:constraint_sdr}, \eqref{eq:s_procedure_constrant}, \eqref{eq:constraint_sen_power_lower_trans}. 
	 	\end{align}
	 \end{subequations}
	Problem \eqref{eq:optimization_problem4} is a convex semidefinite program (SDP) and can be efficiently solved by standard tools such as CVX~\cite{grant2014cvx}.
	Subsequently, the beamforming vector $\mathbf{w}_k$ is recovered from the optimal solution $\mathbf{W}^*_k$ via Gaussian randomization if the rank-one constraint is not satisfied.
	
	\subsection{RA Boresight Optimization}
	
	For fixed $\{\{\mathbf{W}_k\}, \mathbf{R}_{\text{AN}}, \{\lambda_k\}\}$, the RA boresight optimization subproblem can be formulated as 
	\begin{equation}
		\label{eq:optimization_problem5}
			\max_{\boldsymbol{\Theta}_{\text{t}}, \boldsymbol{\Theta}_{\text{r}}} \min_{k} R_k, \  \  
			{\rm{s.t.}} \  \   \eqref{eq:constraint_theta_u},	\eqref{eq:s_procedure_constrant}, \eqref{eq:constraint_sen_power_lower_trans}.
	\end{equation}
	
	Given that RA boresight optimization is highly non-convex with a large search space, conventional convex methods struggle to solve this problem.
	To tackle this challenge, we utilize the PSO algorithm as a suitable alternative.
	We initialize a particle swarm of $S$ particles, with positions $\mathbf{X}^{(0)}$ and velocities $\mathbf{V}^{(0)}$.
	For the $s$-th particle, the initial position is denoted as a vector containing the zenith and azimuth angles for all RAs, i.e., $\mathbf{x}^{(0)}_{s} = \left[\theta^{\text{t}, (0)}_{z, 1}, ... , \theta^{\text{t}, (0)}_{z, N_\text{t}}, \theta^{\text{t}, (0)}_{a, 1}, ... , \theta^{\text{t}, (0)}_{a, N_\text{t}}, \theta^{\text{r}, (0)}_{z, 1}, ... , \theta^{\text{r}, (0)}_{z, N_\text{r}}, \theta^{\text{r}, (0)}_{a, 1}, ... , \theta^{\text{r}, (0)}_{a, N_\text{r}}\right]$.
	
	During the iterative process, each particle is dynamically updated based on the personal best position of the $s$-th particle $\mathbf{x}_{s,\text{p}}$ and the global best position of the entire swarm $\mathbf{x}_\text{g}$.
	In each iteration, the velocity and position of the $s$-th particle are updated according to
	\begin{equation}
		\label{update_v}
		\mathbf{v}^{(l+1)}_{s} = \omega \mathbf{v}^{(l)}_{s} + c_1 r_1 (\mathbf{x}_{s, \text{p}} - \mathbf{x}^{(l)}_s) + c_2 r_2(\mathbf{x}_\text{g} - \mathbf{x}^{(l)}_s),
	\end{equation}
	\begin{equation}
		\label{update_x}
		\mathbf{x}^{(l+1)}_{s} = \mathbf{x}^{(l)}_s + \mathbf{v}^{(l+1)}_{s},
	\end{equation}
	where $l$ denotes the iterative index, $\omega = \omega_{\max} - \frac{\left(\omega_{\max} - \omega_{\min}\right) \cdot l}{L_{\max}}$ is the inertia weight, with bounds $[\omega_{\min}, \omega_{\max}]$  and the maximum iteration $L_{\max}$, $c_1, c_2$ are acceleration coefficients, and $r_1, r_2 \in [0, 1]$ are random variables.
	After each iteration, a boundary check is performed to ensure constraint \eqref{eq:constraint_trans_power}, where any zenith angle exceeding $\theta_{\max}$ is adjusted to the nearest boundary.

	By introducing the penalty factors $\rho_1$ and $\rho_2$, we transform problem \eqref{eq:optimization_problem5} into an unconstrained form.
	The corresponding fitness function is denoted as
	\begin{equation}
		\label{eq:fit}
		\mathcal{F}\left(\mathbf{x}^{(l)}_s\right) = \min_{k} R_k - \rho_1 \mathcal{P}_1 \left(\mathbf{x}^{(l)}_s\right) - \rho_2 \mathcal{P}_2 \left(\mathbf{x}^{(l)}_s\right),
	\end{equation}
	where $\mathcal{P}_1 (\mathbf{x}^{(l)}_s)$ and $\mathcal{P}_2 (\mathbf{x}^{(l)}_s)$ represent the penalty terms, which are given by
	\begin{equation}
		\label{P2X}
		\mathcal{P}_1(\mathbf{x}^{(l)}_s) = \sum_{k=1}^{K} \max(0, -\lambda_{\min}(\mathbf{A}_k)),
	\end{equation}
	\begin{equation}
		\label{P3X}
		\mathcal{P}_2(\mathbf{x}^{(l)}_s) = [\tau - \text{Tr}(\hat{\mathbf{H}}_{\text{R}} \mathbf{R}_{\text{x}} \hat{\mathbf{H}}^H_{\text{R}}) + 2 \varepsilon_H \| \mathbf{R}_{\text{x}} \hat{\mathbf{H}}^H_{\text{R}} \|_F]^+,
	\end{equation}
	where $\lambda_{\min}(\mathbf{A}_k)$ is the minimum eigenvalue for matrix $\mathbf{A}_k$, and $[x]^+ = \max(0, x)$, the penalty terms $\mathcal{P}_1 (\mathbf{x}^{(l)}_s)$ and $\mathcal{P}_2 (\mathbf{x}^{(l)}_s)$ ensure the constraints \eqref{eq:s_procedure_constrant} and \eqref{eq:constraint_sen_power_lower_trans}, respectively.
	
	With the interior-point method, the computational complexity of the SCA algorithm is $\mathcal{O}\left(I_{\text{SCA}} K^{3.5} N^{6.5}\right)$, where $I_{\text{SCA}}$ is the number of SCA iterations. 
	Moreover, the computational complexity of the PSO algorithm is $\mathcal{O}\left(L_{\text{max}} S K N^3\right)$.
	Consequently, the overall computational complexity is	$\mathcal{O}\left(I_{\text{AO}} ((I_{\text{SCA}} K^{3.5} N^{6.5} + L_{\text{max}} S K N^3)\right)$, where $I_{\text{AO}}$ is the total number of AO iterations.
	
	\section{Simulation Results}
	In this section, simulation results are presented to evaluate the performance of our proposed algorithm for the RA-enhanced secure ISAC system.
	We assume that the Eve is located at $[7.5, 3, 13]^T$, and $K=3$ users are located at $[26, -3, 15]^T$, $[21, -3, 21]^T$, and $[-15, -3, 26]^T$, respectively.
	The maximum transmit and noise powers are set to $P_{\max}$ = $30$~dBm and $\sigma_k^2 = \sigma_e^2 = -90$~dBm, repectively.
	For CSI errors, we set $\varepsilon_h = \mu \|\hat{\mathbf{h}}_0\|$ and $\varepsilon_H = \mu \|\hat{\mathbf{H}}_R\|_F$, where $\mu$ is the channel imperfection factor \cite{MengSecure2024}.
    Other parameters are given as follows:
	$p = 3$, $\zeta = 40$~dB, $\ell_u = 2.4$, $\ell_e = 2$, $|\eta| = 1$, $S = 300$, $\omega_\text{max} = 0.9$, $\omega_\text{min} = 0.1$, $c_1 = c_2 = 1.5$, $\rho_1 = \rho_2 = 10^8$, and $L_\text{max} = 300$.
	For performance comparison, we consider the following benchmarks: 1) \textbf{Fixed-orientation scheme}: 
	All RAs are fixed to their initial reference orientations, given by $\boldsymbol{\theta}^{\text{u}}_{n_{\text{u}}} = \mathbf{0}_{2 \times 1}, \forall n_{\text{u}}$.
	2) \textbf{Isotropic antenna scheme}: The antenna elements are isotropic with $p = 0$, where the radiation energy is uniformly distributed within the front half-space.
	
	\begin{figure*}[t]
		\centering
		\subfigure[] { \label{multi_antennas} 
			\includegraphics[width=2.3in]{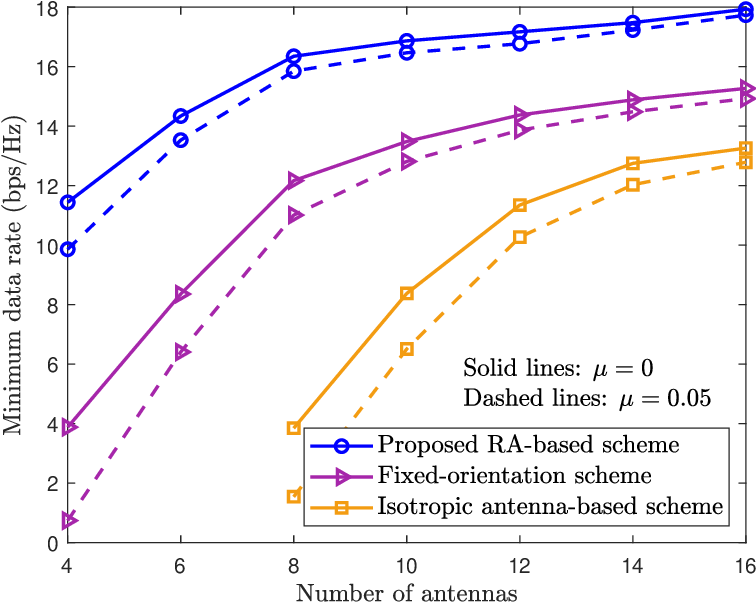} 
		} \hspace{-4mm}
		\subfigure[] { \label{multi_sinr} 
			\includegraphics[width=2.3in]{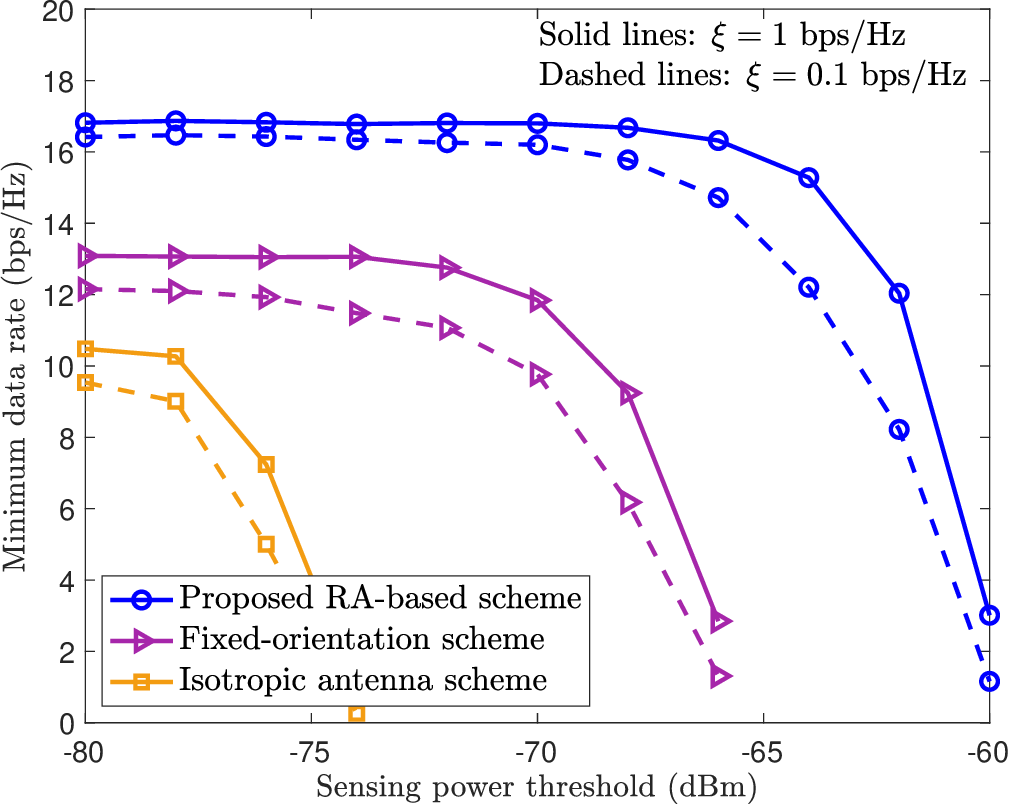} 
		} \hspace{-4mm}
		\subfigure[] { \label{rotatable_range} 
			\includegraphics[width=2.3in]{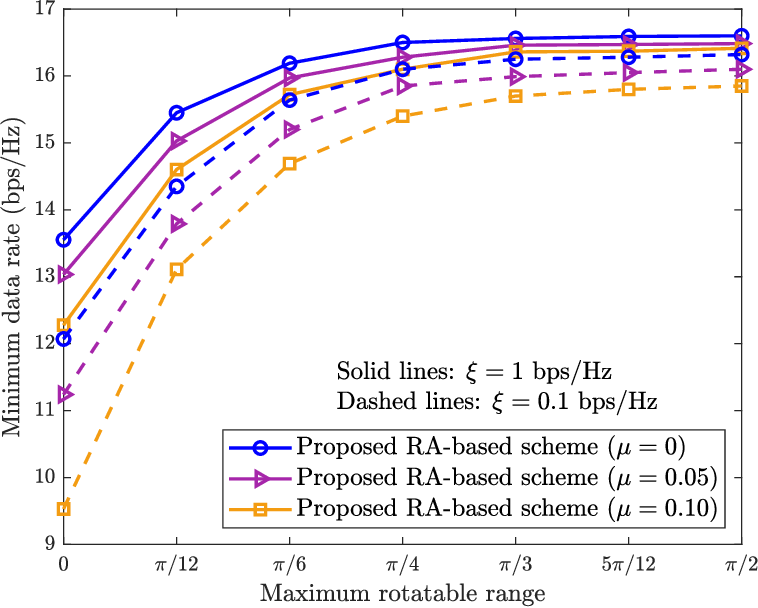} 
		} 
		\caption{{Performance comparisons of the minimum data rate versus (a) the number of antennas $N$; (b) the sensing power threshold $\tau$; (c) the maximum rotation range $\theta_{\max}$.}}
	\end{figure*}
	
	Fig.~\ref{multi_antennas} presents the minimum data rates versus the number of antennas $N$ under both perfect and imperfect CSI scenarios, with $\theta_{\max}=\frac{\pi}{3}$, $\xi_k = \xi = 0.1$~bps/Hz, and $\tau = -75$~dBm.
	As observed, the proposed RA-based scheme significantly outperforms both the fixed-orientation and isotropic antenna schemes. 
	This performance gain stems from the additional spatial DoFs and directional gain brought by RAs, which facilitates enhanced signal strength in desired directions and higher capability for suppressing inter-user interference and information leakage.
	Notably, the proposed RA-based scheme exhibits the least performance loss under the imperfect CSI case, indicating that flexible antenna orientations improve robustness against channel estimation errors.
	Furthermore, as the number of antennas increases, the minimum data rates of all schemes improve, and the performance gap between perfect and imperfect CSI cases gradually narrows.
	This trend is attributed to the expanded array aperture that provides higher beamforming gain and DoFs to mitigate the impact of channel uncertainty.
	
	Fig.~\ref{multi_sinr} depicts the minimum data rate versus the sensing power threshold $\tau$ under various eavesdropping rate constraints $\xi_k=\xi, \forall k$, with $\theta_{\max}=\frac{\pi}{3}$, $N = 8$, and $\mu = 0.05$. 
	It is worth noting that the fixed-orientation and isotropic antenna schemes can enable the maximum sensing power to be up to $-66$~dBm and $-74$~dBm, after which problem (11) becomes infeasible. Meanwhile, the proposed RA-based scheme can sustain a sensing power threshold up to $-60$~dBm.
	This is because benchmarks fail to provide sufficient communication gain under a stringent sensing threshold, while the proposed scheme can enhance the communication and sensing trade-off by dynamically providing higher spatial DoFs.
	It is also observed that the minimum data rate drops at a stricter eavesdropping rate constraint, which is because more power is allocated to the AN to jam the Eve, thus leaving less power for information transmission.
	Nevertheless, the proposed RA-based scheme significantly expands the achievable sensing-communication region compared to the benchmarks, thus demonstrating its effectiveness in the secure ISAC system.
	
	Fig. \ref{rotatable_range} illustrates the minimum data rate versus the maximum rotation range $\theta_{\max}$ under different CSI uncertainty levels and information leakage rate constraints, with $N=8$, $\xi_k=\xi=0.1$~bps/Hz, and $\tau = -75$~dBm.
	We first observe that the minimum data rate improves as the maximum rotation range increases, verifying that the rotational DoFs provided by the RAs can be exploited to effectively reconfigure the channel conditions, thereby enhancing the communication performance.
	Notably, the performance gaps caused by the CSI uncertainty and the information leakage rate constraints gradually narrow as the maximum rotation range increases.
	These observations suggest that the proposed RA-based scheme is more robust against imperfect CSI, and can effectively reduce information leakage, with higher spatial flexibility.
	
	\section{Conclusion}
	This paper investigated an RA-enhanced secure ISAC system, where a transceiver equipped with RAs communicated with users and sensed a target, which was regarded as an Eve. 
	Under the imperfect eavesdropping CSI, we aimed at maximizing the minimum data rate by jointly optimizing the transmit beamforming, AN covariance matrix, and transceiver boresights, while guaranteeing both the minimum sensing power and the maximum information leakage rate. 
	Simulation results demonstrated that the proposed RA-based scheme could provide superior robust minimum data rate performance among users under the given sensing and eavesdropping constraints.
	Moreover, the robustness of the RA-based scheme could be further enhanced with increased rotation range.
	
\bibliographystyle{IEEEtran}
\bibliography{Ref1}

\end{document}